\documentclass{arxiv}
\usepackage{eurovis2026}

\EuroVisShort  

\usepackage[T1]{fontenc}
\usepackage{dfadobe}  

\usepackage{cite}  
\BibtexOrBiblatex
\electronicVersion
\PrintedOrElectronic
\ifpdf \usepackage[pdftex]{graphicx} \pdfcompresslevel=9
\else \usepackage[dvips]{graphicx} \fi

\graphicspath{{./figures/}}

\usepackage{egweblnk}

\title{Unfolding Ordered Matrices into BioFabric Motifs}

\author[J. Wulms, W. Meulemans, \& B. Speckmann]
{\parbox{\textwidth}{\centering J. Wulms\thanks{J. Wulms and W. Meulemans are (partially) supported by the Dutch Research Council (NWO) under project number VI.Vidi.223.137.}\orcid{0000-0002-9314-8260},
        W. Meulemans{$^\dagger$}\orcid{0000-0002-4978-3400},
        and
        B. Speckmann\orcid{0000-0002-8514-7858}
        }
         \\         
{\parbox{\textwidth}{\centering TU Eindhoven, the Netherlands}}%
}

\volume{36}   
\issue{1}     
\pStartPage{1}      

\usepackage{amsmath}
\usepackage{amssymb}
\usepackage{wrapfig}
\usepackage{stfloats}

\usepackage{algorithm}
\usepackage[noend]{algorithmic}

\usepackage{xspace}
\newcommand{\etal}{et~al.\xspace}

\newcommand{\mypar}[1]{\smallskip\noindent\textbf{#1}}

\begin{document}

\teaser{
 \includegraphics[width=0.95\linewidth]{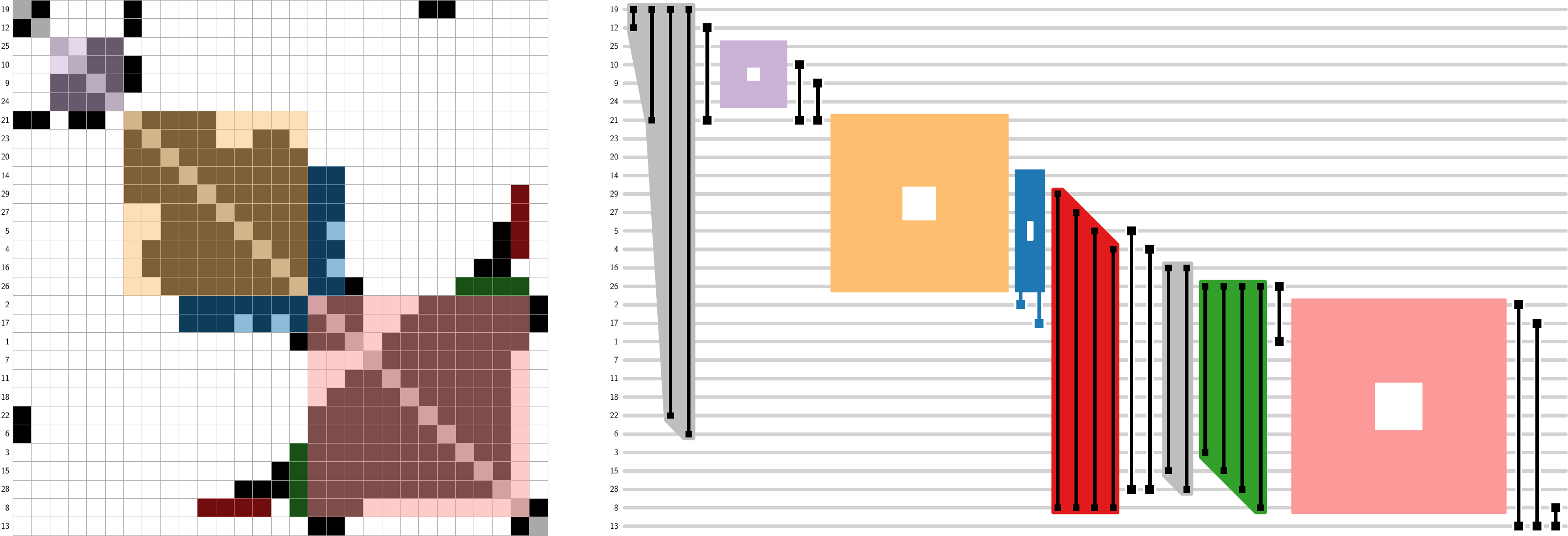}
 \centering
  \caption{Left: a graph represented by a well-ordered (Moran's~$I$) adjacency matrix; our recent algorithm~\cite{DBLP:journals/corr/abs-2601-11171} automatically detects graph patterns (colored rectangles) with $\sigma=0.5$ and $\tau=0.85$. Right: we create a BioFabric motif simplification of the same graph by using the matrix ordering for the vertex lines,``unfolding'' the matrix by traversing it row-by-row, and transforming the patterns into motif glyphs.}
\label{fig:teaser}
}

\maketitle
\begin{abstract}
    BioFabrics were introduced by Longabaugh in 2012 as a way to draw large graphs in a clear and uncluttered manner. 
    The visual quality of BioFabrics crucially depends on the order of vertices and edges, which can be chosen independently. Effective orders can expose salient patterns, which in turn can be summarized by motifs, allowing users to take in complex networks at-a-glance. However, so far there is no efficient layout algorithm that automatically recognizes patterns and delivers both a vertex ordering and an edge ordering to express these patterns as motifs. In this paper we show how to use well-ordered matrices as a tool to efficiently find good vertex and edge orders for BioFabrics. Specifically, we order the adjacency matrix of the input graph using Moran's $I$ and detect (noisy) patterns with our recent algorithm. 
    In this note we show how to ``unfold'' the ordered matrix and its patterns into a high-quality BioFabric. 
    Our pipelines easily handles graphs with up to 250 vertices. 
%
%
%
\end{abstract}  
\section{Introduction}
Visualizing large graphs effectively is a perennial challenge for all that make use of large relational data sets. Most visual designs utilize one of two standard approaches, namely node-link diagrams or adjacency matrices. The BioFabric visualization introduced by Longabaugh~\cite{Longabaugh2012} in 2012 is one of only a few hybrids. A BioFabric represents each vertex with a horizontal line and each edge as vertical connector between its two vertex lines. By design, BioFabrics do not have edge crossings; hence they have the potential to scale better than node-link diagrams while still representing more of the graph topology than adjacency matrices. 

The ``best-of-both-worlds'' design of BioFabrics has recently attracted significant attention. Specifically, Fuchs~\etal~\cite{DBLP:journals/cgf/FuchsDHKB24} explored the design space afforded by the visual separation of vertices and edges into horizontal and vertical lines. The expansion of vertices from zero-dimensional features into one-dimensional lines opens up a host of additional annotation possibilities, especially in combination with the vertically separated edges. 

The visual quality of BioFabrics crucially depends on the order of their vertices and edges, which can be chosen independently. Fuchs~\etal~\cite{BioFabricsQualityMetrics25} cataloged patterns that might be revealed to the user given a well-ordered BioFabric. While the patterns are named according to their visual appearance in a BioFabric, such as runway, staircase, or escalator, they generally correspond to established graph patterns such as cliques, bicliques, or stars. The authors also explored combinations of vertex and edge ordering algorithms and found that generally the edge order influences the detected patterns more strongly. Finding good (edge) orders that reveal patterns effectively was left as an open problem.
Soon after, Di Bartolomeo, Wallinger, and Nöllenburg~\cite{BioFabricsStaircases2025} described an Integer Linear Program (ILP) that computes optimal vertex and edge orderings to reveal staircase (star) patterns in BioFabrics. The results of the ILP showed that optimally ordered graphs reveal more patterns than graphs ordered by heuristic approaches. However, as ILPs are among the more computationally expensive optimization methods, the authors did not compute solutions for graphs with more than 50 vertices.

While BioFabrics can in principle visualize graphs of arbitrary size, they do become unwieldy in their native form for graphs with hundreds, or even thousands, of vertices and edges. Hence Fuchs~\etal~\cite{BioFabricsMotifs2026} recently proposed to use visual motifs to replace patterns and thereby simplify and compress BioFabrics. The authors established that users are generally more effective when working with compressed BioFabrics. However, the presented examples were created based on a simple heuristic ordering with manual fine tuning. An efficient layout algorithm which automatically recognizes patterns, and sorts nodes and edges such that these patterns can be expressed as motifs, was left as an open problem.

\mypar{Contribution.} In this note we show how to effectively ``unfold'' patterns detected in well-ordered matrices into a motif-simplified BioFabric. As mentioned above, Fuchs~\etal~\cite{BioFabricsQualityMetrics25} observed that edge orders have a stronger influence than vertex orders on the patterns that are revealed in a BioFabric. This observation is consistent with our findings. We order the adjacency matrix of the input graph using Moran's $I$, originally a spatial auto-correlation measure, as the quality metric. Van Beusekom, Meulemans, and Speckmann~\cite{DBLP:journals/tvcg/BeusekomMS22} previously established that matrices optimized for Moran's $I$ reveal more visual structure than those ordered via other methods. This is due to the fact that Moran's $I$ promotes edge clusters (clustered black squares in a standard matrix visualization). These edge clusters in turn are exactly the building blocks  both of established graph patterns and of the BioFabric patterns identified by Fuchs~\etal~\cite{BioFabricsQualityMetrics25}.

Our pipeline proceeds as follows. Given an input graph, we represent it as an adjacency matrix and compute an optimal matrix ordering for Moran's $I$. 
Subsequently, we use our recent algorithm~\cite{DBLP:journals/corr/abs-2601-11171} to detect (noisy) graph patterns in the ordered matrix (see Section~\ref{sec:pipeline} for a brief sketch). In Section~\ref{sec:glyphs} we introduce our versions of the BioFabric motifs. We use the matrix order as the vertex order of the BioFabric and explain in Section~\ref{sec:unfold} how to unfold the matrix and its patterns into an ordered set of edges based on a row-by-row traversal. We showcase our results in Section~\ref{sec:case-study}.

\section{Noisy Patterns via Well-Ordered Matrices}
\label{sec:pipeline}

We recently developed an algorithm~\cite{DBLP:journals/corr/abs-2601-11171} that detects noisy patterns in an undirected, unweighted graph. The crucial insight underlying our method is that a well-ordered adjacency matrix allows the high-level graph structures (cliques, bicliques, and stars) to visually emerge as rectangular submatrices. Yet, patterns seldom emerge in perfect form, and thus we account for noise (i.e., missing edges). Our algorithm detects a set of noisy patterns that (1) concisely yet fully capture the high-level graph structure, (2) do not repeat the same structure twice and thus are pairwise disjoint, and (3) the set prioritizes purer patterns over noisier ones.

The algorithm in~\cite{DBLP:journals/corr/abs-2601-11171} proceeds as follows. We first compute an optimal matrix ordering for Moran's $I$. Then we detect submatrices that describe patterns that are dense, structured, and tight. This step is controlled via two parameters $\sigma$ and $\tau$ which capture the purity level of the patterns (values in $[0,1]$; high = more pure, less noise): the ratio of black-black adjacencies between rows/columns should be at least $\sigma$ and the ratio of row/column pairs that adhere to $\sigma$ must be at least $\tau$. Finally, we compute an optimal set of disjoint noisy cliques, complemented with a heuristically detected set of disjoint noisy bicliques and stars. We showcase the results of our algorithm via a so-called Ring Motif simplification, which visually encodes the patterns via circles (cliques) and squares (bicliques and stars), using a hole in these glyphs to encode the level of noise. 

\section{From Noisy Patterns to BioFabric Motifs}
\label{sec:glyphs}

Fuchs~\etal~\cite{BioFabricsMotifs2026} proposed a motif simplification for BioFabrics based on four 
graph patterns: cliques, connectors, fans, and paths. These patterns correspond to four matrix patterns: cliques, bicliques, stars, and bands, respectively. The [paths\,|\,bands] pattern is a weak pattern which does not capture a strong relation between vertices: it generally scores low on  Moran's~$I$~\cite{DBLP:journals/tvcg/BeusekomMS22} and is often part of larger clique or biclique patterns. We hence focus on the first three patterns in our discussion.

A pattern in a matrix is a rectangular submatrix and captures the set of edges present within that submatrix. There are two types of such rectangular submatrices: squares along the diagonal [cliques] and rectangles of any aspect ratio which do not contain the diagonal [bicliques]. [Stars\,|\,fans] are ``skinny bicliques'' of width or height one.
Below we introduce our versions of the BioFabric motifs.

\setlength{\intextsep}{3pt}
\setlength{\columnsep}{9pt}
\begin{wrapfigure}{r}{0.19\columnwidth}
        \includegraphics[width=0.19\columnwidth]{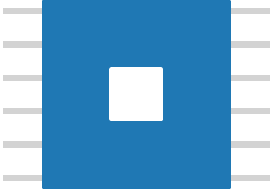}
\end{wrapfigure}
 We use square annulus glyphs to visualize~[cliques]. The outer square intersects the vertex lines of the vertices in the clique. The original BioFabric motif~\cite{BioFabricsMotifs2026} for a clique is a filled square with potentially some missing edges inscribed. However, a square clique represents $O(n^2)$ edges and can accommodate only $n$ inscribed edges, limiting its range. Following our Ring Motif simplification~\cite{DBLP:journals/corr/abs-2601-11171} we rather show a hole proportional to the number of missing edges. As before, a perfect clique is a solid square.

\setlength{\intextsep}{3pt}
\setlength{\columnsep}{9pt}
\begin{wrapfigure}{r}{0.19\columnwidth}
    \centering
    \includegraphics[width=0.19\columnwidth]{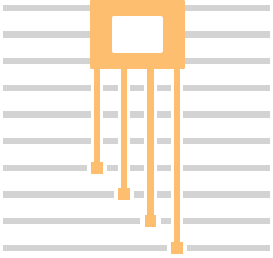}
\end{wrapfigure}
Consider a [biclique] where both sets have at least two vertices (otherwise it is a [star\,|\,fan]).
Let $k$ denote the size of the first vertex set (in vertex order), and $k'$ the size of the second vertex set.
We place a rectangular annulus of width $k'$ over the $k$ (necessarily consecutive) vertex lines corresponding to first set of the biclique. This annulus is connected to the $k'$ vertex lines of the second set using BioFabric edges. As with cliques, the hole of the annulus is proportional to the number of missing edges.

Fuchs~\etal~\cite{BioFabricsMotifs2026} did not introduce a glyph for general bicliques, but their connector pattern is a $(k\times 2)$ biclique. Their glyph is targeted to this special case: it connects two thicker bars with the $k$ vertex lines in between. This glyph does not readily extend to more general bicliques and restricts the possible vertex orderings unnecessarily. In comparison, our design has several advantages: (1) it can encode any biclique (including connectors), (2) it allows us to explicitly encode noise, and (3) the expected vertex ordering coincides with that of a well-ordered matrix. The first two advantages are apparent from the design; we elaborate on the latter advantage in Section~\ref{sec:unfold}.

\setlength{\intextsep}{3pt}
\setlength{\columnsep}{9pt}
\begin{wrapfigure}{r}{0.2\columnwidth}
    \centering
    \includegraphics[width=0.19\columnwidth]{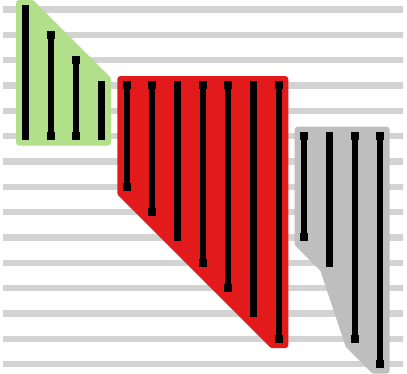}
\end{wrapfigure}
[Stars\,|\,fans] are $(1\times k)$ or $(k\times 1)$ bicliques. Within a BioFabric they are usually called staircase motifs. We follow the previous convention to draw these staircases downwards for $(1\times k)$ or upwards for $(k\times 1)$, ordered by increasing or decreasing edge length, respectively.
Finally, to further declutter the visualization, we aggregate, for each row of the ordered matrix, the edges that are not included in a pattern into a gray downward staircase. 

\begin{figure}[t]
    \centering
    \includegraphics[width=\linewidth]{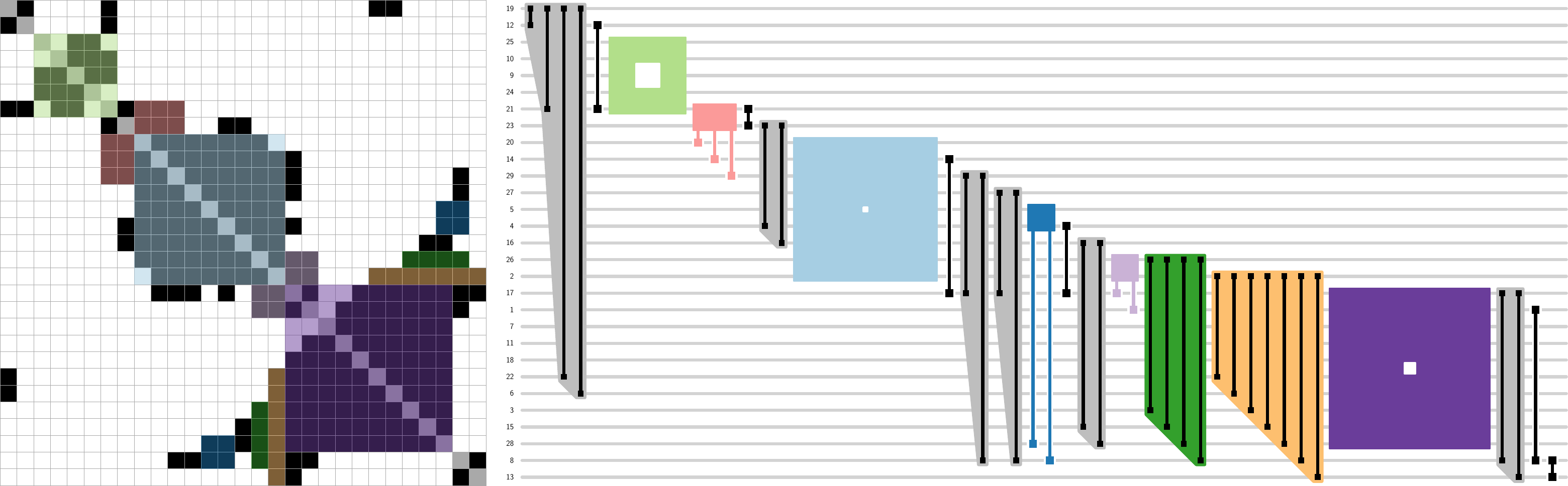}
    \caption{Matrix 58 of the FLT data set with $\sigma = 0.6$ and $\tau = 0.95$.}
    \label{fig:flt-tight}
\end{figure}

\begin{figure*}[b]
    \centering
    \includegraphics[height=110pt]{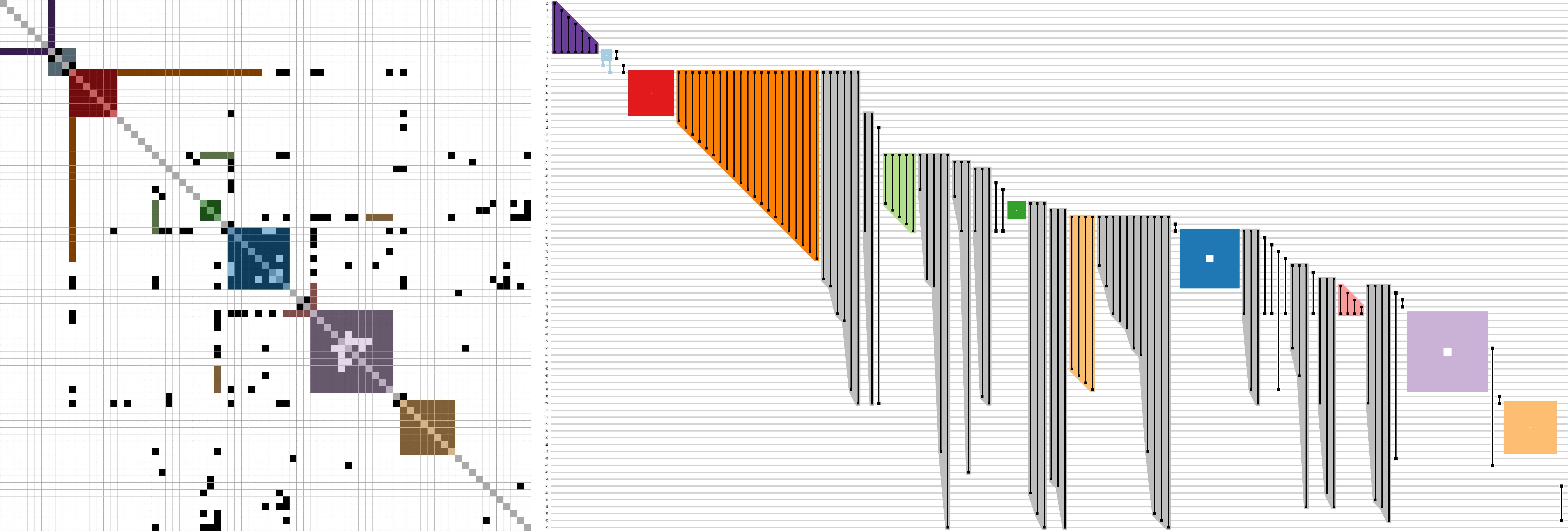}
    \includegraphics[height=110pt]{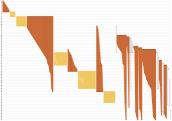}
    \caption{The MIS data set, with vertex ordering proposed in~\cite{BioFabricsMotifs2026} and pipeline parameters $\sigma = 0.6$ and $\tau = 0.96$. The matrix (left) shows identified patterns; our automatically generated motif simplification (middle) finds the same cliques as \cite{BioFabricsMotifs2026} (right).}
    \label{fig:lesmis-motif}
\end{figure*}

\section{Unfolding a Matrix and its Patterns}\label{sec:unfold}
\begin{figure*}[t]
    \centering
    \includegraphics[width=\linewidth]{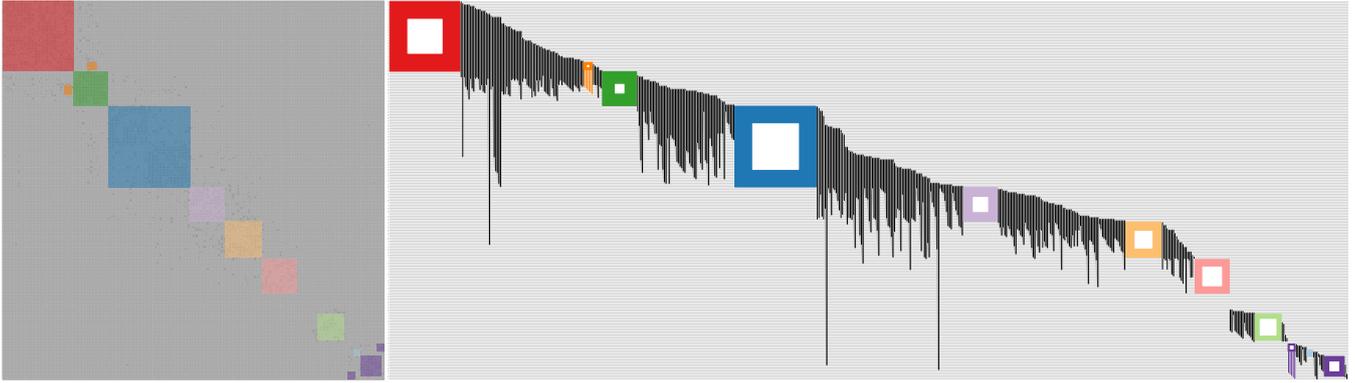}
    \caption{Automated motifs for matrix 2 in the SCH data set with pipeline parameters $\sigma = 0.5$ and $\tau = 0.95$}
    \label{fig:sch02-optimal}
\end{figure*}

We use the matrix order as the vertex order. In the following we describe how to unfold the matrix and its patterns into an ordered set of edges. Every edge has two incident vertices which are assigned an index by the vertex order; we process each edge using its ordered index tuple $(i_1, i_2)$ with $i_1 < i_2$. In the upper part of the matrix (above the diagonal) this tuple corresponds simply to the row index followed by the column index. Each edge is part of at most one pattern and each pattern has a first cell (the upper left corner of the rectangular submatrix that defines the pattern).

The ordering algorithm is now quite simple. We traverse the matrix above the diagonal row by row from left to right and add edges to the order incrementally. When we encounter the first cell of a pattern we add all edges in the pattern consecutively to the order and mark them as ordered. Since we draw motifs, the edge order within a pattern is of no consequence. While traversing a row, we collect all unmarked edges and place them consecutively, once all patterns in the row have been processed. This allows us to further declutter the visualization by aggregating them into the gray ``non-pattern'' staircases we discussed above.

If there are no patterns, and hence all edges are unmarked when we encounter them, then this results in a single staircase for each vertex~$v$. This staircase includes only those edges that connect $v$ to vertices later in the ordering (lower in the matrix). By design, this staircase is top aligned (at $v$) and its edges are increasing in length.

\section{Case Study}\label{sec:case-study}

We demonstrate our automatic procedure for BioFabric motifs simplifications, using four real-life data sets. These data sets arise from different domains and exhibit varying structural properties.

\begin{description}
    \item[FLT:] The ``flashtap'' data set used for MultiPiles~\cite{multipiles}. It is a temporal data set consisting of 96 time steps, each representing functional brain connectivity measured in a Parkinson’s disease study. The graphs each have 29 vertices and are rather dense.
    \item[MIS] The well-known Les Mis\'erables data set~\cite{lesmis}, summarizing character interactions in the Les Mis\'erables book. This graph has 77 vertices, representing characters, which decompose into clear subgroups depending on their roles in the story.
    \item[ZKC:] A social network of karate club members, studied by Zachary~\cite{zachary-karate}. The data set has 34 vertices, representing the members of the club, and the edges model social interactions of club member outside of their club meetings.
    \item[SCH:] Temporal network data representing social interaction between children and teachers at a primary school~\cite{Gemmetto2014, 10.1371/journal.pone.0023176}. The data set is split into 1-hour intervals, resulting in 17 graphs. Each graph has 242 vertices and is fairly sparse.
\end{description}

\mypar{Implementation.}
We build on our previous implementation in Java 21~\cite{DBLP:journals/corr/abs-2601-11171} which is available online~\cite{code}. The matrix reordering step of the pipeline utilizes the Concorde TSP solver~\cite{concorde} on the NEOS server~\cite{neos}. All figures in Section~\ref{sec:case-study}, as well as Figure~\ref{fig:teaser} are produced using this implementation; the pipeline parameter settings are denoted in the caption of each figure.

\subsection{Results}
\begin{figure}[b]
    \centering
    \includegraphics[width=\linewidth]{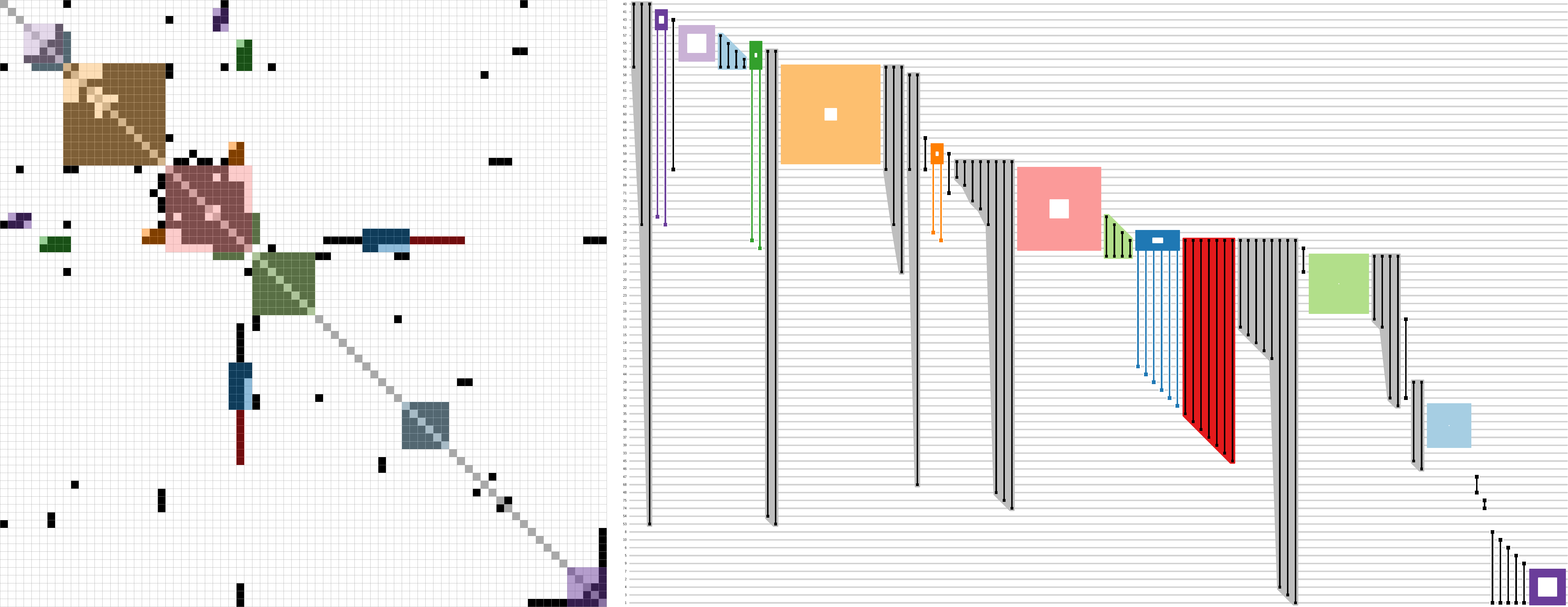}
    \caption{Automated motifs for MIS with $\sigma = 0.3$ and $\tau = 0.9$.}
    \label{fig:lesmis-optimal}
\end{figure}

\mypar{Running times.} Computationally, the matrix reordering step of the pipeline is the most demanding. Even so, the NEOS server finds a TSP tour within 20s, even on the SCH instances. All other parts of the pipeline are executed on a modern laptop (Intel Core i7-12700H, 2300Mhz, 14 Cores, running Windows 11) and finish almost instantaneously: pattern detection usually takes <1s to finish.

\mypar{FLT.} With this dataset we showcase how different pipeline settings result in different BioFabric motifs. Compare Figures~\ref{fig:teaser} and~\ref{fig:flt-tight}: in the former the parameters $\sigma=0.5$ and $\tau=0.85$ were used, while the latter uses higher values and therefore introduces less noise, resulting in more patterns containing less edges.

\mypar{MIS.} The MIS data set is often used to showcase visualizations, for example BioFabric motif simplifications. In~\cite{BioFabricsMotifs2026}, the authors manually tune the vertex and edge orderings to find large cliques and fans. In Figure~\ref{fig:lesmis-motif}, we compare this visualization to a BioFabric made with our pipeline. We enforced the same vertex ordering and let our pipeline detect patterns. We find four of the same cliques (intersecting the same vertex lines) and the fifth clique is a biclique in our case, since it overlaps with another clique; overlapping patterns are disallowed in our pipeline. We also uncover many of the large staircases that show up in the manual ordering of~\cite{BioFabricsMotifs2026}. 
However, Figure~\ref{fig:lesmis-optimal} shows that there are more high-level structures in this data set, which are found by running our full pipeline.

\begin{figure}[b]
    \centering
    \includegraphics[width=0.91\linewidth]{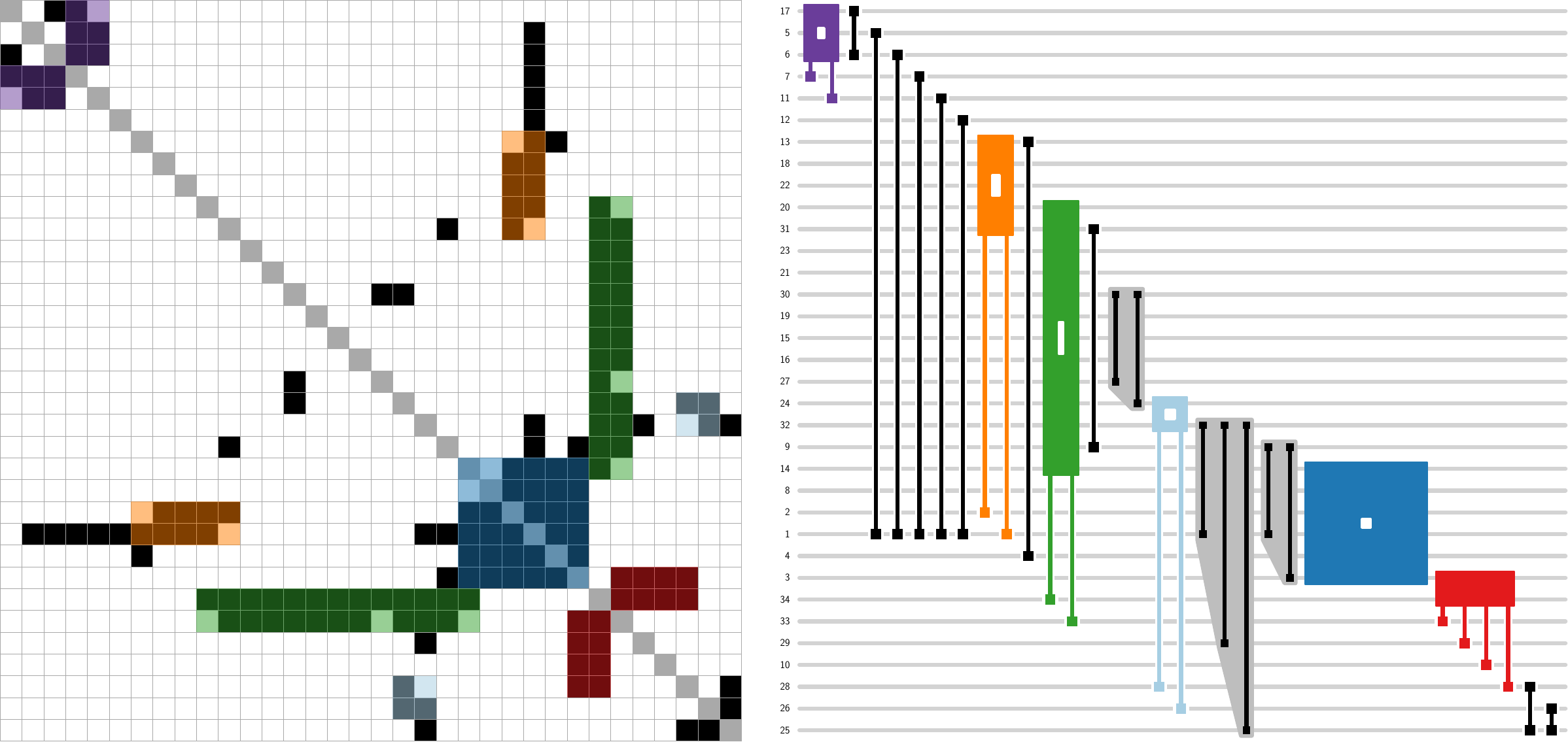}
    \caption{Automated motifs for ZKC with $\sigma = 0.5$ and $\tau = 0.95$}
    \label{fig:karate-optimal}
\end{figure}

\mypar{ZKC.} In Figure~\ref{fig:karate-optimal} we showcase that many bicliques are hidden in the data set. This data set was also used in~\cite{BioFabricsMotifs2026}, but via manual tuning, only one clique and one biclique was found. This again shows the strength of our algorithmic optimization of Moran's~$I$.

\mypar{SCH.} The SCH data set is much larger and therefore a prime target for BioFabric visualizations. In such large graphs, patterns are crucial to find structure. In Figure~\ref{fig:sch02-optimal} we see exactly that: the (bi)clique motifs and the long outlier edges stand out most. Surprisingly, we can also see empty vertex lines, uncovered by Moran's~$I$.

\section{Conclusion}\label{sec:conclusion}
To produce BioFabric motif simplifications, we introduced an automated method that ``unfolds'' a well-ordered matrix and its patterns into a BioFabric. A crucial component of our pipeline is the optimal matrix ordering for Moran's~$I$ to  uncover graph patterns.

\newpage
\bibliographystyle{eg-alpha-doi} 
\bibliography{references}       


\end{document}